\documentclass{aa}
\usepackage{graphicx}
\usepackage{longtable}
\usepackage{txfonts}
\begin{document}
   \title{A long-term photometric study of the FU Orionis star V 733 Cep}


   \author{S. P. Peneva\inst{1}
          \and    
          E. H. Semkov\inst{1}
          \and
          U. Munari\inst{2}
          \and
          K. Birkle\inst{3}}

   \offprints{E. Semkov (esemkov@astro.bas.bg)}

    \institute{Institute of Astronomy, Bulgarian Academy of Sciences,
              72 Tsarigradsko Shose blvd., BG-1784 Sofia, Bulgaria\\
              \email{speneva@astro.bas.bg}
         \and
             INAF Osservatorio Astronomico di Padova, Sede di Asiago, I-36032 Asiago (VI), Italy
         \and
             Max-Planck-Institut f\"{u}r Astronomie, K\"{o}nigstuhl 17, D-69117 Heidelberg, Germany}

   \date{Received ; accepted }


  \abstract
   {The FU Orionis candidate V733 Cep was discovered by Roger Persson in 2004. 
   The star is located in the dark cloud L1216 close to the Cepheus OB3 association. 
   Because only a small number of FU Orionis stars have been detected to date, photometric and spectral studies of V733 Cep are of great interest.}
   {The studies of the photometrical variability of PMS stars are very important to the understanding of stellar evolution. 
   The main purpose of our study is to construct a long-time light curve of V733 Cep.
   On the basis of $BVRI$ monitoring we also study the photometric behavior of the star.}
   {We gather data from CCD photometry and archival photographic plates.
   The photometric $BVRI$ data (Johnson-Cousins system) that we present were collected from June 2008 to October 2009.
   To facilitate transformation from instrumental measurements to the standard system, fifteen comparison stars in the field of V733 Cep were calibrated in $BVRI$ bands.
   To construct a historical light curve of V733 Cep, a search for archival photographic observations in the Wide-Field Plate Database was performed.
   As a result, 192 photographic plates containing the field of V733 Cep were found.
   Some plates were analyzed at our request to estimate the magnitude of V733 Cep.}
   {Our photometric study confirms the affiliation of V733 Cep to the group of FU Orionis objects. 
   An outburst in the optical and a slow rise in brightness during the period 1971-1993 are well documented.
   During the period 1993-2004, V733 Cep exhibited its maximum brightness and the amplitude of the observed outburst exceeded 4$\fm$5 (R). 
   The $BVRI$ photometric data imply that from February 2007 to October 2009, a slow decrease in brightness of V733 Cep was observed.
   The observed color evolution of $V-I$ index also suggest that V733 Cep is currently fading.
   The long-time light curve of V733 Cep is similar to the light curves of other FU Orionis objects.}
   {}


   \maketitle
%

\section{Introduction}

During the Pre-Main Sequence (PMS) stage of evolution, young stellar objects exhibit different types of photometric variability (Herbst et al. 1994).
One of the most dramatic of these events, with very high amplitude variations, is the FU Orionis (FUor) outburst (Ambartsumian 1971).
The flare-up of FU Orionis itself was documented by Wachmann (1939) and for several decades it was the only known object of that type.
Herbig (1977) defined FUors as a class of young variables after the discovery of two new FUor objects, V1057 Cyg and V1515 Cyg.
The main characteristics of FUors are an increase in optical brightness of about 4-5 mag, a F-G supergiant spectrum with broad blue-shifted Balmer lines, strong infrared excess, connection with reflection nebulae, and location in star-forming regions (Reipurth 1990, Bell et al. 1995, Clarke et al. 2005).
Typical spectroscopic properties of FUors include a gradual change in the spectrum from earlier to later spectral type from the blue to the infrared, a strong Li I 6707 line, P Cygni profiles of H$\alpha$ and Na I 5890/5896 lines, and the presence of CO bands in the near infrared spectra (Herbig 1977, Bastian \& Mundt 1985).
The prototypes of FUors seem to be low-mass PMS objects (T Tauri stars) with massive circumstellar disks. 

According to a commonly accepted view, the FUor outburst is produced by a sizable increase in accretion from a circumstellar disk on the stellar surface (Hartmann \& Kenyon 1985).
The cause of this increase in accretion from  $\sim$10$^{-7}$$M_{\sun}$$/$yr to $\sim$10$^{-4}$$M_{\sun}$$/$yr appears to be thermal or gravitational instability in the circumstellar disk. 
This accretion disk model can account for most of the main properties of FUors. 
An alternative explanation of the FUor phenomenon is the rapid-rotator hypothesis (Herbig at al. 2003).

Among all objects associated with the group of FUors, only three (FU Ori, V1057 Cyg, and V1515 Cyg) have
detailed photometric observations taken during the outburst and during the fading period (Clarke et al. 2005). 
For a few objects, V1735 Cyg (Elias 1978, Peneva et al. 2009), V346 Nor (Graham \& Frogel 1985), and V733 Cep
(Reipurth et al. 2007), the presence of an optical outburst is also documented and they are labeled classical FUors. 
Two new suspected FUor objects with observed outbursts at optical wavelengths are reported: V582 Aurigae (Samus 2009, Munari et al. 2009) and the object CSS091110, coincident with the infrared source IRAS 06068-0641 (Wils et al. 2009).
About a dozen objects have spectroscopic properties similar to the classical FUors, but there is no evidence of an outburst at optical wavelengths. 
These objects are termed FUor-like (Reipurth et al. 2002, Greene et al. 2008) and only partial photometric observations have been published for them.

The PMS star V733 Cep (Persson's star) is located in the dark cloud L1216 close to Cepheus OB3 association. 
The variability of V733 Cep was discovered by Swedish amateur astronomer Roger Persson in 2004 (Persson 2004). 
Persson compared the plate scans from the first and the second Palomar Sky Survey. 
He noted the star of red magnitude 15$\fm$7 on the POSS-II image (UT 1991 September 3) and its absence on the corresponding POSS-I image (UT 1953 October 31).
The star is also visible on a Palomar Quick-V plate from 1984. 
On an image taken with the 2.2-m telescope in Mauna Kea, Hawaii on UT 2004 October 9, Reipurth et al. (2007) measured a red magnitude of V733 Cep at about 17$\fm$3. 
Comparing this value with that reported by Persson (2004), Reipurth et al. (2007) concluded that the star had faded by 1$\fm$6 (R) over a time period of about 13 yr (from 1991 to 2004).
The authors propose that an outburst occurred in the period 1953-1984 and identify great spectral similarities to FU Ori itself.

In our first paper (Semkov \& Peneva 2008; hereafter SP08), data from photometric monitoring of V733 Cep in the period February 2007 - February 2008 were presented, and it was noted that no significant changes in the star brightness were registered during the period of observations.
Because of the short time period of observations, a trend toward a decrease in brightness was not detected.

\section{Observations}

We present photometric data acquired as part of our continued investigation of V733 Cep (SP08).
The photometric $\it BVRI$ data were collected from June 2008 to October 2009.
Our observations were performed at two observatories with three telescopes: the 2-m Ritchey-Chretien-Coude and 50/70/172 cm Schmidt telescopes of the National Astronomical Observatory Rozhen (Bulgaria) and the 1.3-m Ritchey-Chretien telescope of the Skinakas Observatory\footnote{Skinakas Observatory is a collaborative project of the University of Crete, the Foundation for Research and Technology - Hellas,
and the Max-Planck-Institut f\"{u}r Extraterrestrische Physik.} of the Institute of Astronomy, University of Crete (Greece).
We used the ANDOR CCD camera with the 1.3-m RC telescope, the VersArray CCD camera with 2-m RCC telescope, and a SBIG ST11000 camera
with the 50/70 cm Schmidt telescope.
The technical parameters for the CCD cameras used, observational procedure, and data reduction process are described in SP08.

The standard stars used for comparison are of great importance to the correct magnitude estimation. 
In regions of star formation such as the Cepheus OB3 association, a great percentage of stars can be photometric variables. 
In SP08, we presented $BVRI$ photometric data (Johnson-Cousins system) for fifteen stars in the vicinity of V733 Cep suitable for comparison. 
Using new photometric data we try to improve the $BVRI$ magnitudes of the comparison stars and to eliminate possible low amplitude variable stars. 
Calibrations were made with the 1.3 m RC telescope during two clear nights in July and August 2009. 
Standard stars from Landolt (1992) were used as a reference. 
Table 1 contains our corrected photometric data for the $BVRI$ comparison sequence. 
The corresponding mean errors in the mean are also listed. 
The stars are labeled from A to O in order of their V-band magnitude.
The finding chart of the comparison sequence and the coordinates of the stars are presented in SP08. 
Our new photometric data of the comparisons agree well with the published ones in SP08 and the corresponding mean errors were improved considerably. 
The standards range from $V$ = 15$\fm$052 to $V$ = 19$\fm$697 and from $V-I$ = 1$\fm$805 (star B) to $V-I$ = 4$\fm$883 (star F). 
The availability of such a large number of standard stars is necessary for a correct measuring of the archival photographic observations.
The photographic emulsion is a non-linear receiver and measured object need to be compared with stars of similar magnitude.
We suspect that the stars C, F, and K from our list are possible variables of small amplitude and we advise observers to use these data with care.

\begin{table*}
\caption{Photometric data for $BVRI$ comparison sequence}
\label{table:2}
\centering
\begin{tabular}{lllllllll}
\hline\hline
\noalign{\smallskip}
Star &  $V$ & $\sigma_V$& $Ic$ & $\sigma_I$ & $Rc$ & $\sigma_R$ & $B$ & $\sigma_B$ \\
\noalign{\smallskip}
\hline
\noalign{\smallskip}
A	&	15.052	&	0.037	&	12.570	&	0.041	&	13.782	&	0.026	&	17.160	&	0.051	\\
B	&	16.084	&	0.032	&	14.279	&	0.031	&	15.124	&	0.028	&	17.569	&	0.039	\\
C	&	16.749	&	0.030	&	12.595	&	0.077	&	14.614	&	0.043	&	19.962	&	0.124	\\
D	&	17.080	&	0.031	&	14.440	&	0.039	&	15.747	&	0.035	&	19.241	&	0.087	\\
E	&	17.090	&	0.030	&	14.362	&	0.036	&	15.729	&	0.035	&	19.226	&	0.080	\\
F	&	17.508	&	0.045	&	12.625	&	0.071	&	15.166	&	0.060	&	20.535	&	0.176	\\
G	&	17.718	&	0.038	&	14.627	&	0.041	&	16.152	&	0.034	&	20.221	&	0.176	\\
H	&	17.740	&	0.038	&	15.541	&	0.033	&	16.653	&	0.046	&	19.498	&	0.095	\\
I	&	17.810	&	0.045	&	14.996	&	0.036	&	16.440	&	0.041	&	19.940	&	0.052	\\
J	&	17.954	&	0.055	&	16.037	&	0.038	&	17.006	&	0.035	&	19.500	&	0.085	\\
K	&	18.102	&	0.084	&	14.950	&	0.059	&	16.606	&	0.054	&	20.386	&	0.203	\\
L	&	18.140	&	0.045	&	14.983	&	0.042	&	16.539	&	0.038	&	20.633	&	0.251	\\
M	&	18.325	&	0.048	&	15.743	&	0.037	&	17.135	&	0.035	&	20.164	&	0.110	\\
N	&	18.618	&	0.048	&	16.714	&	0.046	&	17.666	&	0.026	&	20.174	&	0.092	\\
O	&	19.697	&	0.134	&	16.680	&	0.062	&	18.185	&	0.067	&	$-$	&	$-$	\\
\hline                                   
\end{tabular}
\end{table*}

The results of our photometric observations of V733 Cep are summarized in Table 2.
The columns indicate Julian date of observation, $\it IRVB $ magnitudes, and telescope used.
The typical instrumental errors from CCD photometry are in the range $0\fm01$-$0\fm02$ for $I$ and $R$, $0\fm03$-$0\fm05$ for $V$ and $0\fm05-0\fm11$ for $B$ filter.
The $\it BVRI$ light curves of V733 Cep during the period of our observations (SP08 and the present paper) are plotted in Fig. 1.

   \begin{table*}
   \centering
   \small
   \caption[]{$\it BVRI $ CCD photometric observations of V733 Cep}
   \begin{tabular}{llllllllllll}
            \hline \hline
            \noalign{\smallskip}
      JD (24...) & $\it I$  & $\it R$ & $\it V$ & $\it B$  &  Telescope & JD (24...) & $\it I$  & $\it R$ & $\it V$ & $\it B$  &  Telescope\\
            \noalign{\smallskip}
            \hline
            \noalign{\smallskip}
54632.579	&	14.21	&	16.45	&	18.30	&	-	    &	1.3-m	  &	54996.570	&	14.27	&	16.52	&	-   	&	-	    &	1.3-m	\\
54638.418	&	14.24	&	16.49	&	18.58	&	-	    &	1.3-m	  &	55000.564	&	14.28	&	16.55	&	18.57	&	-	    &	1.3-m	\\
54646.524	&	14.17	&	16.48	&	18.48	&	21.16	&	1.3-m	  &	55003.575	&	14.27	&	16.57	&	18.61	&	-	    &	1.3-m	\\
54647.503	&	14.22	&	16.51	&	18.50	&	21.17	&	1.3-m	  &	55006.567	&	14.37	&	16.74	&	18.80	&	21.14	&	1.3-m	\\
54653.461	&	14.15	&	16.44	&	18.42	&	21.00	&	1.3-m	  &	55009.540	&	14.44	&	16.83	&	18.92	&	21.40	&	1.3-m	\\
54654.494	&	14.17	&	16.44	&	18.43	&	20.99	&	1.3-m	  &	55014.543	&	14.37	&	16.69	&	18.77	&	21.49	&	1.3-m	\\
54656.441	&	14.26	&	16.52	&	18.50	&	-	    &	1.3-m	  &	55016.545	&	14.32	&	16.59	&	18.62	&	-	    &	1.3-m	\\
54658.570	&	14.20	&	16.50	&	18.50	&	-	    &	1.3-m	  &	55022.560	&	14.52	&	16.81	&	18.87	&	-	    &	1.3-m	\\
54660.588	&	14.27	&	-    	&	-	    &	-	    &	1.3-m	  &	55027.543	&	14.62	&	16.97	&	-	    &	-	    &	Schmidt	\\
54661.470	&	14.28	&	16.67	&	18.73	&	21.30	&	1.3-m	  &	55028.515	&	14.66	&	16.98	&	-	    &	-	    &	Schmidt	\\
54672.495	&	14.21	&	16.55	&	18.54	&	-	    &	1.3-m	  &	55029.535	&	14.67	&	16.98	&	-	    &	-	    &	Schmidt	\\
54673.530	&	14.23	&	16.54	&	18.59	&	21.61	&	1.3-m	  &	55031.584	&	14.66	&	16.94	&	18.98	&	-	    &	1.3-m	\\
54680.556	&	14.15	&	16.40	&	18.39	&	-	    &	1.3-m	  &	55037.560	&	14.49	&	16.73	&	18.67	&	-	    &	1.3-m	\\
54706.432	&	14.25	&	16.21	&	-   	&	-	    &	Schmidt	&	55041.358	&	14.40	&	16.65	&	18.66	&	-	    &	1.3-m	\\
54707.449	&	14.28	&	16.26	&	18.31	&	-	    &	Schmidt	&	55044.511	&	14.18	&	16.43	&	18.43	&	20.90	&	1.3-m	\\
54761.291	&	14.22	&	16.35	&	-   	&	-	    &	Schmidt	&	55048.580	&	14.22	&	16.50	&	-	    &	-	    &	1.3-m	\\
54763.273	&	14.18	&	16.37	&	18.43	&	-	    &	Schmidt	&	55065.403	&	14.26	&	16.56	&	-	    &	-	    &	Schmidt	\\
54791.280	&	14.32	&	16.46	&	-	    &	-	    &	Schmidt	&	55066.306	&	14.31	&	16.56	&	-	    &	-	    &	Schmidt	\\
54842.239	&	14.27	&	16.43	&	-	    &	-	    &	Schmidt	&	55111.455	&	14.34	&	16.59	&	-	    &	-	    &	Schmidt	\\
54844.218	&	14.31	&	16.55	&	-	    &	-	    &	Schmidt	&	55112.447	&	14.30	&	16.55	&	-	    &	-	    &	Schmidt	\\
54917.552	&	14.52	&	16.77	&	-	    &	-	    &	Schmidt	&	55113.416	&	14.30	&	16.52	&	-	    &	-	    &	Schmidt	\\
54938.742	&	14.93	&	17.10	&	-	    &	-	    &	Schmidt	&	55114.314	&	14.33	&	16.58	&	-	    &	-	    &	Schmidt	\\
54971.445	&	14.27	&	16.39	&	-	    &	-	    &	Schmidt	&	55128.291	&	14.38	&	-   	&	18.49	&	-	    &	2-m	\\
54994.560	&	14.37	&	16.62	&	18.82	&	-	    &	1.3-m	  &	55133.221	&	14.40	&	16.66	&	-	    &	-	    &	Schmidt	\\
\noalign{\smallskip}
\hline
\end{tabular}
\end{table*}

   \begin{figure}
   \centering
   \includegraphics[width=\columnwidth]{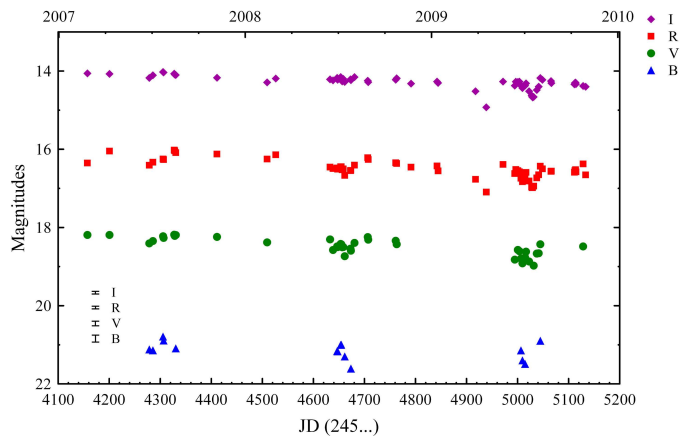}
   \caption{$BVRI$ light curves of V733 Cep. Typical error bars for each filter are shown to the left.}
    \end{figure}

\section{Search for archival photographic observations and measurements of photographic plates}

The construction of the historical light curves of FUor objects would be very important for
the determination of the outburst mechanism.
The only one possible way to study the long-term variability of PMS stars is the photographic plate archives.
The search for archival observations of V733 Cep in the photographic plate collections all over the world was made using the Wide-Field Plate Database (WFPDB) (Tsvetkov et al. 1997).

The WFPDB\footnote{http://vizier.u-strasbg.fr/viz-bin/VizieR, Catalogue VI/90} contains (1) a catalogue of all known
archives of wide-field ($\geq$1$^{\circ}$) plates and (2) a merged catalogue of wide-field plates.
Our search in the database was limited to {\bf clear telescope apertures} $\geq40$ cm.
The results of our detailed examination of the catalogue are summarized in Table 3.
We found 192 archival photographic observations in the plate collections of five telescopes.
In addition, one of us (K. B.) found two plates obtained with the 80/120 cm Schmidt telescope and the 123 cm reflector of
the Max-Planck-Institut f\"{u}r Astronomy, Heidelberg, at Calar Alto Observatory
(the plate catalogues of this telescopes are still not included in the WFPDB).
The first results from the plate archives of Asiago and Calar Alto observatories are reported in the present paper.
The digitized plates from the Palomar Schmidt telescope, available via the website of the
Space Telescope Science Institute, are also used.
The plates obtained with the Kiso Schmidt telescope are presently inaccessible, because of technical problems with the scanning machine at the Kiso observatory. 
We hope to report results from this archival plate collection in a future study.

\begin{table*}
\caption{Archival photographic plates of V733 Cep found in WFPDB.}
\label{table:2}
\centering
\begin{tabular}{lrlrlr}
\hline\hline
\noalign{\smallskip}
Observatory & Telescope & Telescope & Number  & Band & Range \\
            & Aperture (cm) & Type & of plates & pass &of years \\
\hline\noalign {\smallskip}
Harvard & 61 & Rfr. & 5 & pg & 1894-1895 \\
Harvard & 41 & Rfr. & 60 & pg & 1921-1942 \\
Palomar & 122 & Sch. & 9 & pgBVRI & 1953-1993 \\
Asiago  & 67  & Sch. & 40 & BVRI & 1971-1978 \\
Kiso    & 105 & Sch. & 78 & BVRI & 1976-1987 \\
\noalign {\smallskip}  
\hline
\noalign {\smallskip}
Total &      &   & 192 & &1894-1993 \\
\hline                                   
\end{tabular}
\end{table*}

The old plates from the archives of the 61-cm and 41-cm reflectors at the Harvard observatory were obtained using less sensitive emulsions and the plate limit is very low despite the long exposure times.
According to our CCD observations the $B$ magnitude of the star at maximum light does not exceeds 20 mag.
The star is seen just above the limit on the deep blue plates from the Second Palomar Schmidt telescope survey and we do not expect to detect V733 Cep on $B/pg$ plates from telescopes with smaller aperture.
The plates from Calar Alto Observatory were scanned at Max-Planck-Institut f\"{u}r Astronomy, Heidelberg with 2540 dpi resolution, which corresponds to 10$\times$10 $\mu$m pixel size. 
Aperture photometry of the digitized plates from the Palomar Schmidt telescope was performed with DAOPHOT routines. The $BVRI$ comparison sequence
reported in the present paper was used as a reference.

The plates from Asiago Schmidt telescope are inspected visually using a high quality Carl Zeiss microscope, which offers a variety of magnifications (Munari et al. 2001). 
The magnitude is then derived by comparing the stars in the photometric sequence with the variable, identifying those that match most closely the variable.
If ``a" and ``b" were two stars in the sequence, by visual inspection we would estimate the quantities n1 and n2 that represent the fraction of the total a-b magnitude difference by which the variable V is fainter than ``a" and brighter than ``b", i.e. a - n1 - V - n2 - b. 
The magnitude of V follows from simple proportion. 
If more than one such pair were available, more estimates would be derived and weighted for the ``a-b", ``c-d" etc. mag interval.
Typical estimated errors are of the order of 0.10 mag.

The results of estimating magnitudes of the archival photographic plates are summarized in Table 4.
In Fig. 2, we plot the $B$, $V$, $R$, and $I$ light curves for all available observations of V733 Cep. 
The filled triangles denote our CCD observations, the filled diamonds photographic data from the Asiago Schmidt telescope, the filled circles photographic data from the Palomar Schmidt telescope, the open diamonds photographic data from the Calar Alto Observatory, and the open triangle the CCD observation from the 2.2-m telescope in Mauna Kea, Hawaii. 
The arrow indicate the upper limit to the POSS-I red plate magnitude.

\begin{table*}
\caption[]{Photometric data from the photographic observations of V 733 Cep}
\begin{tabular}{lllllr}
              \hline
             \hline
\noalign {\smallskip}  
Observatory & Plate No. & Bandpass & Date  & JD (24...) & Magnitude  \\
            \hline
\noalign {\smallskip}  
Palomar	&	874E  &	$R$	&	1953 Oct. 31	&	34681.597	&	$>$20.5\\
Palomar & 874O  & $pg$ &1953 Oct. 31  & 34681.632 & $>$21.0\\
Asiago  & 4577  & $B$ & 1971 Aug. 16  & 41180.420 & $>$19.5\\
Asiago	&	4578	&	$R$	&	1971 Aug. 16	&	41180.447	&		18.0$\pm$0.1\\
Asiago	&	4579	&	$I$	&	1971 Aug. 16	&	41180.473	&		16.3$\pm$0.1\\
Asiago  & 4669  & $B$ & 1971 Sep. 21  & 41216.503 & $>$ 19.5\\
Asiago	&	4670	&	$I$	&	1971 Sep. 21	&	41216.530	&	$>$	15.7\\
Asiago  & 4709  & $B$ & 1971 Oct. 15  & 41240.351 & $>$ 17.7\\
Asiago	&	4710	&	$I$	&	1971 Oct. 15	&	41240.382	&	$>$	15.7\\
Asiago  & 4810  & $B$ & 1971 Oct. 21  & 41246.338 & $>$ 19.5\\
Asiago	&	4811	&	$I$	&	1971 Oct. 21	&	41246.366	&	$>$	15.5\\
Asiago  & 4989  & $B$ & 1971 Oct. 26  & 41251.519 & $>$ 17.7\\
Asiago	&	5011	&	$I$	&	1971 Dec. 07	&	41293.212	&	$>$	15.5\\
Asiago	&	5012  & $B$ & 1971 Dec. 07  & 41293.236 & $>$ 17.7\\
Asiago	&	5064	&	$I$	&	1971 Dec. 12	&	41298.214	&	$>$	14.3\\
Asiago	&	5065  & $B$ & 1971 Dec. 12  & 41298.241 & $>$ 17.7\\
Asiago	&	5096	&	$I$	&	1971 Dec. 15	&	41301.224	&		15.7$\pm$0.1\\
Asiago	&	5097  & $B$ & 1971 Dec. 15  & 41301.251 & $>$ 17.7\\
Asiago	&	5656  & $B$ & 1972 Oct. 05  & 41596.458 & $>$ 19.5\\
Asiago	&	5657	&	$I$	&	1972 Oct. 05	&	41596.480	&		16.1$\pm$0.1\\
Asiago	&	5748	&	$I$	&	1972 Nov. 02	&	41624.351	&		16.2$\pm$0.1\\
Asiago	&	5749  & $B$ & 1972 Nov. 02  & 41624.372 & $>$ 19.5\\
Asiago	&	5808	&	$I$	&	1972 Nov. 06	&	41628.338	&		15.7$\pm$0.1\\
Asiago	&	5809  & $B$ & 1972 Nov. 06  & 41628.359 & $>$ 19.5\\
Asiago	&	5934	&	$I$	&	1972 Nov. 29	&	41651.220	&		15.7$\pm$0.1\\
Asiago	&	5935  & $B$ & 1972 Nov. 29  & 41651.242 & $>$ 19.5\\
Asiago	&	5980	&	$I$	&	1972 Dec. 11	&	41663.316	&		15.8$\pm$0.1\\
Asiago	&	5982	&	$V$	&	1972 Dec. 11	&	41663.389	&		18.6$\pm$0.1\\
Asiago	&	6025	&	$I$	&	1972 Dec. 24	&	41676.211	&		15.6$\pm$0.1\\
Asiago	&	6026	&	$V$	&	1972 Dec. 24	&	41676.231	&		18.9$\pm$0.1\\
Asiago	&	9284  & $B$ & 1877 Nov. 07  & 43455.433 & $>$ 17.7\\
Asiago	&	9285	&	$I$	&	1977 Nov. 07	&	43455.458	&		15.3$\pm$0.1\\
Asiago	&	9296	&	$I$	&	1977 Nov. 10	&	43458.349	&		15.2$\pm$0.1\\
Asiago	&	9325	&	$I$	&	1977 Nov. 16	&	43464.379	&		15.2$\pm$0.1\\
Asiago	&	9348  & $B$ & 1977 Dec. 02  & 43480.381 & $>$ 17.7\\
Asiago	&	9349	&	$I$	&	1977 Dec. 02	&	43480.406	&		15.2$\pm$0.1\\
Asiago	&	9374  & $B$ & 1977 Dec. 14  & 43492.356 & $>$ 17.7\\
Asiago	&	9713  & $B$ & 1978 Nov. 01  & 43814.347 & $>$ 19.5\\
Asiago	&	9714	&	$I$	&	1978 Nov. 01	&	43814.374	&		15.1$\pm$0.1\\
Asiago	&	9795	&	$V$	&	1978 Nov. 24	&	43837.360	&	$>$	18.6\\
Asiago	&	9796  & $B$ & 1978 Nov. 24  & 43837.377 & $>$ 17.7\\
Calar Alto	&	A1111	&	$R$	&	1980 Aug. 18	&	44469.518	&		16.9$\pm$0.1\\
Calar Alto	&	E1521	&	$R$	&	1981 Dec. 05	&	44944.323	&		16.9$\pm$0.1\\
Palomar	&	889V	&	$V$	&	1984 Aug. 27	&	45939.859	&		17.75$\pm$0.1\\
Palomar	&	2671	&	$B$	&	1989 Sep. 02	&	47771.776	&		20.9$\pm$0.2\\
Palomar & 4103  & $I$ & 1991 Jul. 23  & 48460.918 &   14.35$\pm$0.05\\
Palomar & 4146  & $B$ & 1991 Aug. 08  & 48476.875 &   20.7$\pm$0.1\\
Palomar	&	4183	&	$R$	&	1991 Sep. 03	&	48502.787	&		16.18$\pm$0.05\\
Palomar & 4191  & $R$ & 1991 Sep. 06  & 48505.802 &   16.17$\pm$0.04\\
Palomar	&	5364	&	$I$	&	1993 Aug. 16	&	49215.805	&		14.22$\pm$0.06\\
\hline
\end{tabular}
\end{table*}

   \begin{figure*}
   \centering
   \includegraphics[width=15cm]{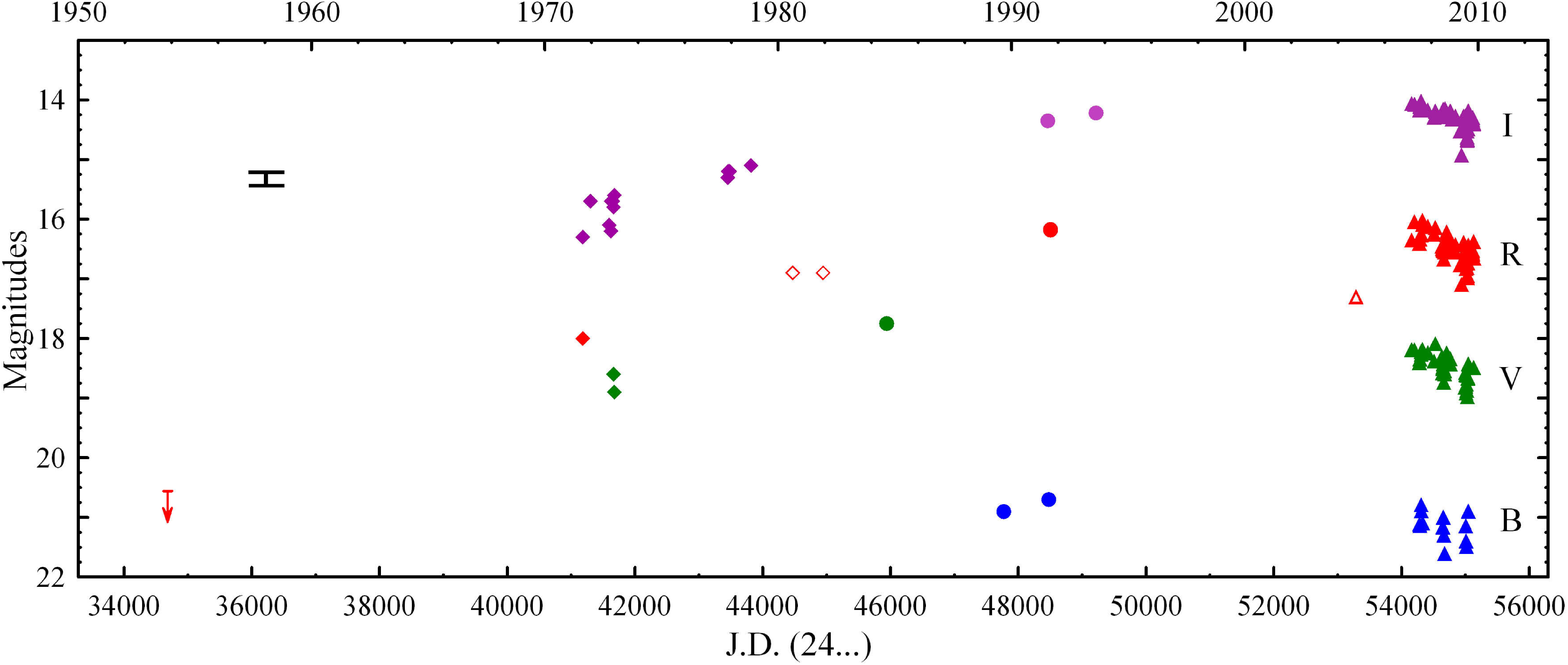}
      \caption{Historical $BVRI$ curves of V 733 Cep. A typical error bar for photographic observtions is shown to the left.}
         \label{Fig2}
  \end{figure*}

\section{Results and discussion}

Our photometric study confirms the affiliation of V733 Cep to the group of FUor objects.
An outburst in optical light and a rise in brightness in the period 1971-1993 are well documented.
During this period, the increase in brightness passed very slowly, and for 22 years the $I$-band magnitude increased by 2$\fm$1.
It is currently impossible to determine the time of the beginning of the optical outburst due to a lack of observations for the period 1953-1971.
The light curve of V733 Cep in the period of increase in brightness is similar to that observed for the FUor star V1515 Cyg, but the time of rise seems to be longer.
During the period 1993-2004, V733 Cep probably reached its maximum brightness for which there are no published photometric observations.
The amplitude of the observed outburst of V733 Cep exceeds 4$\fm$5 ($R$). 

Our CCD photometric data (SP08 and the present paper) imply that from February 2007 to October 2009, a slow decrease in brightness of V733 Cep was observed (Fig. 1).
The observational data infer that the star decreased in brightness in $I$ band by 0$\fm$15 per year, and in $V$-band by 0$\fm$23 per year.  
Another important result from our photometric study is the variation in color indices with stellar brightness. 
In Fig. 3, we plot the measured color index $V - I$ versus stellar magnitude $V$ during the period of our observations. 
A clear dependence can be seen in the period of set in brightness: the star becomes redder as it fades.
According to Hartmann \& Kenyon (1996), this color evolution is observed in the case of FUor stars that have a relatively fast set in brightness (V1057 Cyg).
The explanation of the observed photometric color evolution is that the spectral type of V1057 Cyg changes from early A-type near maximum light (Herbig 1977) to middle G-type at the present (Herbig 2009).
In spite of this, FUor objects that experience a very slow decrease in brightness (V1515 Cyg) do not exhibit this color evolution (Clarke et al. 2005).

The decrease in brightness of V733 Cep proceeds irregularly, as periods of short drops in brightness are observed.
A typical example of this drop in brightness is observed during June-July 2009 (decrease by 0$\fm$4 ($I$) and return to its previous level) (Fig. 1).
During the period of set in brightness, the change in the photometric activity seems to be typical of FUor objects.
A strong decrease in brightness by about $1\fm5$ ($B$) in a few months was registered in the light curve of V1515 Cyg (Clark et al. 2005). 
This minimum in brightness was attributed to obscuration caused by dust material ejected from the star (Kenyon et al. 1991). 
Evidence of strong light variability at the time of set in brightness ($\Delta$$V$=1$\fm$2) were reported in the photometric study of another FUor object -- V1735 Cyg (Peneva et al. 2009).
In 2004, the measured relatively faint red magnitude for V733 Cep (Reipurth et al. 2007) can be attributed similarly to variable obscuration by dust material.

The photometric data that has been collected to date for V733 Cep are insufficient to separate the effects of two suspected causes of reddening: the change in the spectral type and obscuration by a dust material. 
The observed variations in the $VRI$ magnitudes during the three years of photometric monitoring are only a few tenths. 
To understand why the star reddened during the fading period, we plan to acquire new photometric and spectral observations.

The shape of observed light curves of FUors may vary considerably from object to object. 
While the time of rise for FU Ori and V1057 Cyg is approximately 1 year, it is considerably longer for V1515 Cyg ($\sim25$ years).  
The rate of decrease in brightness is also correspondingly different: while the brightness of V1057 Cyg reaches the pre-outburst level after $\sim30$ years, the decrease in brightness of FU Ori, V1735 Cyg, and V1515 Cyg proceeds much more slowly.
The available photometric data for V733 Cep are insufficient at present to determine precisely the amplitude of outburst and the time of rise to the maximum brightness.  
Our conclusion is that the light curve of V733 Cep during the period of rise in brightness is more similar to the light curve of V1515 Cyg,
and during the period of set in brightness is far more similar to the light curve of V1057 Cyg.
Therefore, V733 Cep is probably the first FUor object with approximately symmetrical light curve (the time of rise is similar of the time of setting in brightness).

Because only a small number of objects have been classified as FUors, their classification is very difficult. 
In some cases, stars originally classified as FUors in detailed spectral and photometric studies were identified as variables of other types. 
An example is V1184 Tau (CB 34V), which was initially considered to be a FUor, but is now more likely to be a star of UX Ori type (Barsunova et al. 2006, Semkov et al. 2008).
According to Grinin et al. (2009), these two types of PMS variability (FUor and UX Ori) can be observed for the same object in different periods of time.
Photometric data accumulated over time provides an opportunity to say that the variability of V733 Cep is from FUor type.

\begin{figure}
\centering
\includegraphics[width=\columnwidth]{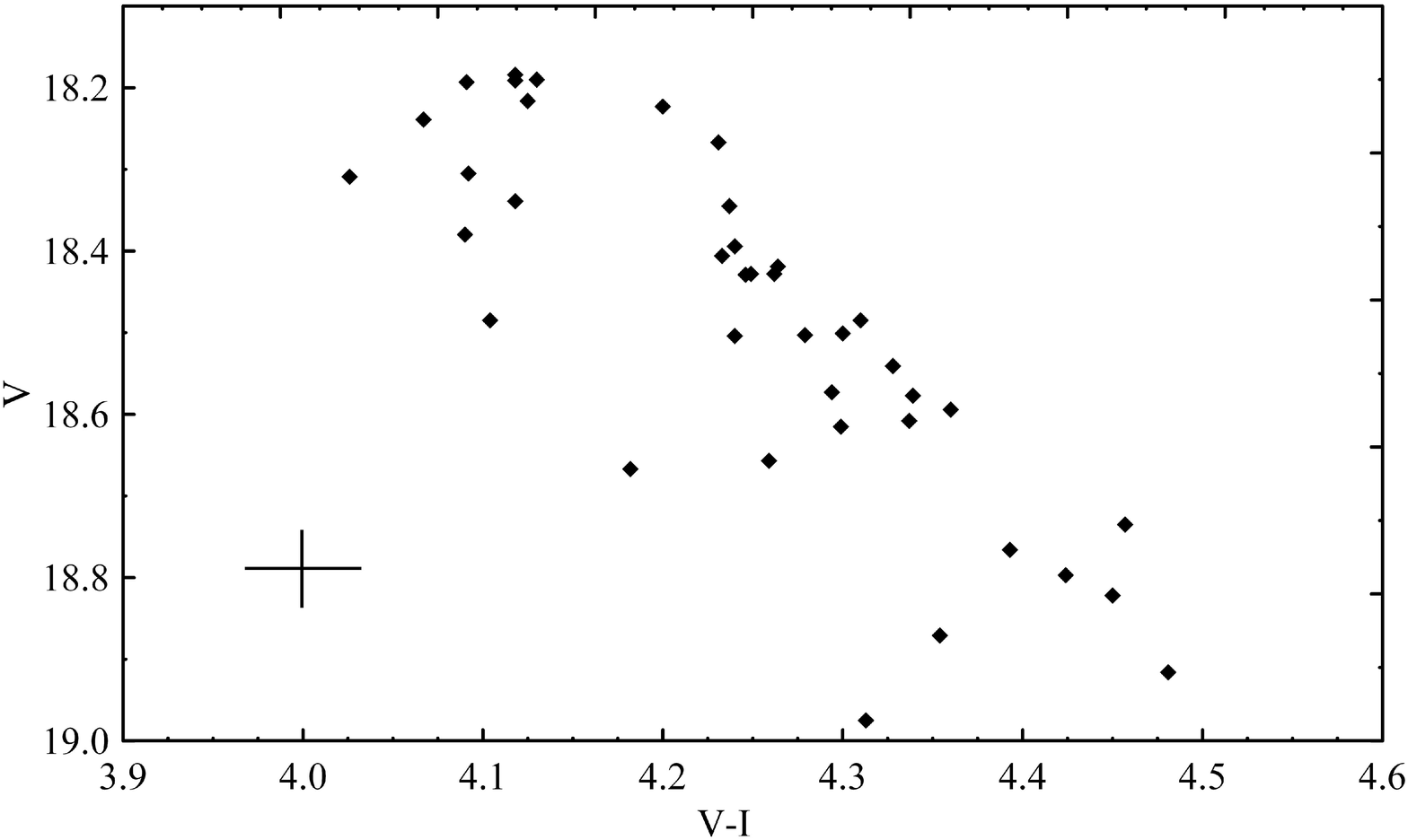}
\caption{$V/V-I$ diagram of V733 Cep in the period Feb. 2007 - Oct. 2009. Typical error bars for the plotted quantities are shown at the left.}
\end{figure}

\section{Conclusions}

The analysis of the available photometric data has allowed us to precisely classify V733 Cep as a FUor variable. 
The long-time light curve of the star is similar to the light curves of others FUor objects. 
The observed photometric variations in the period of fading are also typical of some FUor stars.
The time of rise in brightness and the star magnitude in the maximum light remain unclear. 
Therefore, the collection of new photometric data (from photographic plate archives and from ongoing photometric monitoring) will be of great importance for a precise determination of the outburst parameters.
We conclude that V733 Cep is presently the FUor object with the longest time of increase in brightness and probably the first found to have an approximately symmetrical light curve.

\begin{acknowledgements}
      This work was partly supported by grants DO 02-85, DO 02-273, DO 02-363 and F-201/2006 of the National Science Fund of
      the Ministry of Education, Youth and Science, Bulgaria.
      The authors thank the Director of Skinakas Observatory Prof. I. Papamastorakis
      and Prof. I. Papadakis for the telescope time.
      The Digitized Sky Survey was produced at the Space Telescope Science Institute under
      U.S. Government grant NAG W-2166.
      The images of these surveys are based on photographic data obtained using
      the Oschin Schmidt Telescope on Palomar Mountain and the UK Schmidt Telescope.
      The plates were processed into the present compressed digital form with the permission of these institutions.
      This research has made use of the NASA Astrophysics Data System.
\end{acknowledgements}

\end{document}